
\magnification=1000
\tolerance=10000
\baselineskip=24truept
\def\ref{\par\noindent\hangindent 20pt}

\def\mincir{\raise -2.truept\hbox{\rlap{\hbox{$\sim$}}\raise5.truept
\hbox{$<$}\ }}
\def\magcir{\raise -2.truept\hbox{\rlap{\hbox{$\sim$}}\raise5.truept
\hbox{$>$}\ }}
\def\gr{\kern 2pt\hbox{}^\circ{\kern -2pt K}} 

\def\asymp{\raise -4.3truept\hbox{$ \ \widetilde{\phantom{xy}} \ $}}

\null
\bigskip
\centerline{\bf BLUE PERTURBATION SPECTRA FROM INFLATION}
\bigskip
\bigskip
\centerline{\bf Silvia Mollerach}
\medskip
\centerline{Theory Division, CERN,}
\medskip
\centerline{CH-1211, Geneve 23, Switzerland}
\bigskip
\centerline{\bf Sabino Matarrese}
\medskip
\centerline{Dipartimento di Fisica {\it Galileo Galilei}, Universit\`a
di Padova,}
\medskip
\centerline{via Marzolo 8, I--35131 Padova, Italy}
\bigskip
\centerline{\bf Francesco Lucchin}
\medskip
\centerline{Dipartimento di Astronomia, Universit\`a di Padova,}
\medskip
\centerline{vicolo dell'Osservatorio 5, I--35122 Padova, Italy}
\bigskip
\baselineskip=18truept
\noindent
{\bf Abstract}
\bigskip
We investigate inflationary models leading to density perturbations
with a spectral index $n>1$ (``blue spectra"). These perturbation
spectra may be useful to simultaneously account for both the amount of
ultra large-scale power required to fit cosmic microwave background
anisotropies, such as those measured by COBE, and that required to
give bulk motions and structures on the $\sim 50~h^{-1}$ Mpc scale.
\bigskip
\noindent
{\bf Subject headings:}
\bigskip
Cosmology: theory -- early universe --galaxies: formation --
large-scale structure of the universe.
\vfill
\eject

\noindent
{\bf 1. Introduction}
\medskip

The recent detection by COBE (Smoot {\it et al.} 1992) of anisotropies
in the Cosmic Microwave Background Radiation (CMBR) at large angular
scales has provided an insight on the primordial density
perturbations. This measurement makes it possible to normalize the
power spectrum at the largest observable scales and constrains the
primordial spectral index in the range $n = 1.1^{+0.45}_{-0.55}$. This
is consistent with the well-known Harrison--Zeldovich spectrum
($n=1$), widely used in models of structure formation, but also with a
significant range of values for the spectral index around it. It is
however clear that, to further constrain the shape of the power
spectrum, other measurements at smaller scales are required. One
proposal (Davis {\it et al.} 1992) is to combine the COBE detection
with measurements of the CMBR at the degree angular scale, as tested
by the South Pole experiment (Gaier {\it et al.} 1992; Schuster {\it
et al.} 1993). It has been claimed that these measurements are
consistent with the COBE data for a scale-invariant power spectrum and
the Cold Dark Matter (CDM) scenario; but the present accuracy and the
inherent cosmic variance of the observations do not allow a precise
determination of $n$ (Dodelson \& Jubas 1993; Bunn  {\it et al.} 1993;
see however Gorski, Stompor \& Juszkiewicz 1993).

On the other hand the standard $n=1$ CDM model, normalized at ultra
large scales to COBE ($\sim 10^3h^{-1}$ Mpc), predicts too much power
on scales of the order of $1h^{-1}$ Mpc (where $h$ is the Hubble
constant in units of $100$ km s$^{-1}$ Mpc$^{-1}$), which is in
disagreement with observations such as the pairwise galaxy velocity
dispersion (Davis \& Peebles 1983). Tilted (i.e. $n<1$) CDM models
have been proposed to alleviate this conflict (Vittorio, Matarrese
\& Lucchin 1988; Cen {\it et al.} 1992; Lucchin, Matarrese \&
Mollerach 1992; Adams {\it et al.} 1992; Tormen {\it et al.} 1993).
Other alternatives are based on either invoking a relic cosmological
constant (Efstathiou, Sutherland \& Maddox 1990; Turner 1991; Kofman,
Gnedin \& Bahcall 1993) or adding a hot dark matter component to the
CDM scenario, leading to the so-called mixed dark matter scenarios
(Davis, Summers \& Schlegel 1992; Klypin {\it et al.} 1992). The
latter possibility seems to be among the most promising alternatives
to standard CDM, although at the expense of the introduction of one
more free parameter in the theory: the relative amounts of the two
dark matter components.

Piran {\it et al.} (1993) have recently suggested an alternative view,
which is  to combine the CMBR measurement at large scales and the
observation of large voids in the galaxy distribution. Since the voids
can be assumed to result from underdensities in the true matter
distribution, they can be used to normalize the power spectrum on the
typical scale of the voids detected in the CfA survey (De Lapparent,
Geller \& Huchra 1986; Geller \& Huchra 1989), i.e. of order
$50h^{-1}$ Mpc. This is independent of the galaxy to matter
distribution relation, i.e. of bias. This analysis seems to favour a
spectral index $n \simeq 1.25$. An $n > 1$ (``blue") spectrum,
normalized to COBE, might also provide the large power shown by galaxy
peculiar velocities on similar scales, but it exacerbates the problem
of the scale-invariant spectrum: the excess power on the Mpc scale.
However, as pointed out by Piran {\it et al.} (1993) the extra power
on small scales implies that non-linearity should be important on
scales larger than usually thought, which must be properly taken into
account when studying this range of scales. Another possible way out
of this small scale problem is to invoke a hot dark matter component,
which would wash out the excess small scale power by free streaming.
Moreover, the possible presence of a gravitational wave contribution
to the COBE detection (Lucchin {\it et al.} 1992; Davis {\it et al.}
1992; Lidsey \& Coles 1992; Salopek 1992; Liddle \& Lyth 1992,
Souradeep \& Sahni 1992) allows a further shift to the blue side of
the spectrum, if the above $\approx 50h^{-1}$ Mpc normalization is
taken at its face value. This is because a stochastic background of
gravitational waves only affects the CMBR isotropy, through the
Sachs--Wolfe effect, on scales above $\approx 1^\circ$, i.e. above
$100h^{-1}$ Mpc.

On the other hand it is also possible to put upper bounds on $n$. A
recent study (Carr and Lidsey 1993) has put a limit on the spectral
index, $n<1.5$, from primordial black hole overproduction constraints
on nucleosynthesis and the COBE observations.

Power spectra with $n>1$ have not been very widely studied in the
literature, one reason being that they are not the most common
prediction of theories for the origin of the primordial fluctuations.
Our aim in this paper is to explore inflationary models that originate
this kind of power spectra. We first investigate in the next section
models that can give rise to an exact power-law spectrum putting
emphasis on the  $n>1$ case, and then in section 3 some more realistic
models where an approximate power-law with $n>1$ arises. Finally we
present our conclusions in section 4.

\medskip
\noindent
{\bf 2. Inflationary models with exact power-law spectra}
\medskip

Quantum fluctuations of scalar fields during inflation are one of the
most appealing explanations for the origin of the primordial density
perturbations.  An inflationary period in the early universe gives
rise to both scalar and tensor perturbations in the metric. The scalar
perturbations, associated to fluctuations in the energy density, are
often described in terms of the gauge-invariant variable $\zeta$. Its
power spectrum  to first order in the slow roll approximation is given
by
$$P_{\zeta} ^{1/2} (k) = \left. { H^2  \over 2 \pi \dot \phi}
\right|_{aH = k}. \eqno (1) $$

For the gravitational wave spectrum we have
$$ P_{GW} ^{1/2} (k) =\left. { H\over 2 \pi} \right|_{aH = k}.
\eqno (2) $$

Chaotic  inflationary models based on a polynomial potential  give
rise to a power spectrum $P_\zeta ^{1/2} (k)$ which decreases with $k$
but only logarithmically. Another example is the power-law inflation,
which is based on an exponential potential for the inflaton; it gives
rise to a power spectrum decreasing with $k$ as a power, $P_\zeta
^{1/2} (k) \propto k^ \alpha$, with $\alpha <0 $ (Lucchin \& Matarrese
1985). Power spectra
decreasing with $k$ are the most usual outcome of inflationary models
involving one minimally coupled scalar field. This is determined by
the dependence of $H^2/\vert\dot \phi\vert$ on $\dot a$ in equation (1).
Or, equivalently, as by definition during inflation $\dot a$ increases
with time ($\ddot a > 0$), we can say that $d {(H^2/\vert\dot
\phi\vert)} /dt < 0$ gives rise to a power spectrum decreasing with $k$
and $d {(H^2/\vert\dot \phi\vert)}/ dt > 0$ to one increasing with $k$.
The fact that the time dependence of $H^2$ is usually the dominant one
and $H^2$ is always decreasing with time makes $P_\zeta $ decrease
with $k$. However, this is not a necessary conclusion: inflationary
models in which $\vert\dot \phi \vert^{-1}$ increases with time more
rapidly than $H^2$ decreases will originate a power spectrum $P_\zeta$
increasing with $k$. We want to analyse the possibility that a power
spectrum increasing with $k$ as a power arises during the inflationary
dynamics, that is $P_\zeta^{1/2} (k) \propto k^\alpha$ with $\alpha >
0$. This corresponds to a spectral index $n \equiv 2 \alpha + 1 > 1$
for the density perturbations on constant time hypersurfaces before
matter-radiation equality.

According to equation (1) a power spectrum $P_\zeta ^{1/2} \propto
k^\alpha$ results when
$$ {H^4 \over \dot \phi ^2}
= A (aH) ^{2 \alpha}, \eqno (3) $$
with $ A = H_*^4 /(\dot {\phi}_*^2
a_*^{2\alpha} H_*^{2\alpha}) $, where the subscript $*$ indicates
that the quantities are
evaluated at the time a given wave-number $k_*$ crossed the horizon
$k_* = a_* H_*$. Using the equation of motion
$ 2\dot H = - \kappa^2 \dot \phi ^2$,
with $\kappa^2 \equiv8 \pi/ M_P^2$, we obtain
$$ \dot  H = -{ \kappa^2 H^{4-2\alpha} \over 2
A a^{2\alpha}}. \eqno (4) $$

Using the expansion parameter $a$ as a
``time" variable we can write equation (4) as
$$ H' = -{ \kappa^2
H^{3-2\alpha} \over 2 A a^{2\alpha + 1}}, \eqno (5) $$
where $' \equiv d/da$, which can be easily integrated.

We first consider the case $0 \ne \alpha \ne 1$. From equation (5) we
get
$$ H^2 = \left( {\kappa^2 \over 2A} {\alpha -1 \over \alpha} a^{-2
\alpha} + B \right)^{1/(\alpha - 1)}, \eqno (6) $$
with
$$ B = -{\kappa ^2 \over 2A} {\alpha -1 \over \alpha} a_0^{-2 \alpha}
+  H_0^{2(\alpha -1)}. \eqno (7) $$
In the $\alpha = 1$ case  we get
$$ H =
H_0 \exp \left( {\kappa^2 \over 4 A} \left( {1 \over a^2} - {1
\over a_0^2} \right) \right), \eqno (8) $$
and for $\alpha = 0$
$$H^{-2} = H_0^{-2} + { \kappa^2 \over A } \ln \left( { a \over a_0}
\right). \eqno (9) $$

We can now obtain $\dot \phi^2 (a)$ by replacing $H(a)$ from equations
(6), (8) or (9) in equation (3). Using again $a$ as variable we can
integrate it to obtain $\phi( a)$
$$ d \phi = \pm da { H^{1-\alpha}
(a) \over \sqrt {A} a ^{1 + \alpha}}. \eqno (10) $$
For the three  cases $0 \ne \alpha \ne 1, \alpha =1, \alpha =0$, the
integration can be analytically performed, and the result inverted to
obtain $a(\phi)$. With all these elements it is  possible to determine
the inflaton potential that originates these spectra. Replacing
$\dot\phi^2 (a)$ from equation (10) and $H(a)$ from equations (6), (8)
or (9) into $$ V = {3 H^2 \over \kappa^2} - {\dot \phi ^2 \over 2},
\eqno (11)$$ we obtain $V(a)$ and, as we also have $a(\phi)$, we can
obtain the inflaton potential $V(\phi)$. Since the computations depend
on the value of $\alpha$ considered, they have to be analysed
separately.

For $0 \ne \alpha \ne 1$ the integration of equation (10) with $H(a)$
given by equation (6) depends on the sign of $(\alpha - 1) / \alpha$
and $B$. We are going to analyse in detail the case $(\alpha - 1) /
\alpha  < 0$ and then outline the main results for the other cases.
The considered case corresponds to values of $\alpha$ in the range $ 0
< \alpha < 1$, or to values of $n$ in the range $ 1 < n < 3$. From
equation (7) we see that in this case $B> 0$. Integrating equation
(10) and inverting the resulting expression one easily obtains
$$ a^{2 \alpha} = {\kappa^2 (1 - \alpha)
\over 2 \alpha AB} \cos ^{-2} (\psi), \eqno (12) $$
with $ \psi \equiv\phi \sqrt {{1 \over 2} \kappa^2 \vert \alpha (1 -
\alpha) \vert } $. Replacing in equation (11) equations (3), (6) and
(12), the resulting potential is:
$$
V(\phi) = {B^{{1 \over \alpha - 1}} \over \kappa^2 (1 - \alpha)} \sin
^{2 (2 - \alpha)/( \alpha - 1)} ( \psi) \left( -\alpha +(3 - 2 \alpha )
\sin ^2 (\psi) \right). \eqno  (13) $$
This potential has a minimum at $ \psi = \pi /2$, two maxima, one to
the right and the other to the left of the minimum, and it goes to $-
\infty$ for $\psi$ equal to $0$ and $\pi$. The potential minimum is
$V_{min} = 3B^{1 \over \alpha - 1} / \kappa^2$. (While writing this
paper, we learnt that a similar solution has been obtained by Carr and
Lidsey, 1993.)

The  solution  obtained is inflationary, i.e. $\ddot a > 0$, only when
$\dot \phi ^2 < V(\phi)$. Replacing equations (6) and (12) into
equation (3) and comparing with equation (13) one can show that
inflation occurs only for values
of $\psi$ such that $\sin^2 (\psi) > \alpha$, that is in a region
around the potential minimum compressed between the two maxima. Thus,
inflation will occur while the inflaton field rolls down to its
potential minimum, starting when its kinetic energy becomes small
enough at values of $\phi$ such that $\sin ^2 (\psi) > \alpha$. From
equation (12) we see that infinite e-foldings occur befor the inflaton
arrives at the minimum: thus the field does not oscillate around the
minimum and the model has no built-in reheating mechanism. As in the
exponential potential inflation we have to invoke another process to
end inflation.

It is interesting to analyse the contribution of the gravitational
wave modes in this model. According to Lucchin {\it et al.}
(1992), the relative contribution of the tensor and scalar
modes to the cosmic microwave background radiation anisotropies is
given by
$$
{(\delta T/T)_S \over  (\delta T /T)_T} \simeq {2 \over 5} {H \over
\kappa \dot \phi}. \eqno (14) $$
In our case we obtain
$$ {(\delta
T/T)_S \over  (\delta T /T)_T}  = { 2 \over 25} {1 - \alpha  \over
\alpha} \tan^2 (\psi). \eqno (15) $$
We see that for values  of $\phi$ close to $\sin ^2 (\psi) = \alpha$,
where inflation starts, $(\delta T/T)_S / (\delta T /T)_T \sim  2 / 25
$, thus the gravitational wave contribution can dominate;  however, as
$\phi$ evolves towards the potential minimum, the contribution becomes
always smaller. We can compute the number of e-foldings during which
tensor modes dominate.  From equation (12) one sees that the number of
e-foldings it takes for $\phi$ to evolve from $\phi_x$ to $\phi_y$ is
given by
$$ \ln \left({ a_y \over a_x}\right) = {1 \over 2 \alpha }
{\cos ^2 (\psi_x) \over \cos ^2 ( \psi_y)}. \eqno (16) $$
Taking $\phi_y$ as the value of $\phi$ for which the scalar and tensor
contributions are equal, we have from equation (15) $\cos^2 \psi_y = 2
(1 - \alpha ) /( 23 \alpha + 2)$, and $\phi_x$ as the starting value
$\cos^2 \psi_x = 1 - \alpha$, we get
$$\ln \left({a_y \over a_x} \right) = {1 \over 2 \alpha} \ln \left(1 +
{23 \over 2} \alpha \right). \eqno (17) $$
Thus, the gravitational wave contribution to the CMBR can only be
dominant for the first few e-foldings of inflation.

The other cases $(\alpha - 1)/\alpha > 0$, $\alpha = 0$ and $\alpha =
1$ can be solved following the same steps. We are going only to
outline the main results.

For $(\alpha -1)/ \alpha > 0$ ( $\alpha < 0$ or $ \alpha > 1$) we see
from equation (9) that the sign of the constant $B$ is no longer fixed
and we have to consider the three possibilities: $B = 0$, $ B > 0$ and
$B < 0$. The first  case leads  to an exponential potential $V(\phi)
\propto \exp ( \pm \kappa \phi\sqrt{2 \alpha/ (\alpha - 1)} )$ and
corresponds to the well-known power-law inflation case. Inflation can
only happen for $\alpha < 0$. The universe expands by $\ln (a_y/a_x) =
\kappa \vert\phi_x - \phi_y\vert \sqrt {(\alpha - 1)/2 \alpha } $
e-foldings when the field evolves from $\phi_x$ to $\phi_y$. The
relative contribution of scalar and tensor modes to the CMBR is
$(\delta T/T)_S^2/(\delta T/T)_T^2 = 2(\alpha -1)/25 \alpha$. We are
now going to quote : a) the potential, b) the condition for inflation
$(\ddot a > 0)$, c) the number of e-foldings it takes the field to
evolve  between two values, d) the ratio of the scalar to tensor
contribution to CMBR, and will discuss the main consequences for the
other cases.

\smallskip \noindent Case $B>0$
$$\eqalignno{{\rm a)}& \ \ \ \ V(\psi) = {B^{1 \over \alpha - 1} \over
\kappa^2 (1 - \alpha)} \cosh^{2(2-\alpha)/(\alpha - 1)} (\psi) \left(
- \alpha + (3 - 2\alpha ) \cosh^2(\psi) \right) & (18a) \cr
{\rm b)} & \ \ \ \  \tanh^2 (\psi) < (\alpha -1) / \alpha  \to \ddot a
>0 & (18b) \cr
{\rm c)} & \ \ \ \  \ln \left({a_y \over a_x}\right) = {1 \over 2
\alpha} \ln \left({\sinh^2 (\psi_x) \over \sinh ^2 (\psi_y)} \right) &
(18c) \cr
{\rm d)} & \ \ \ \ {(\delta T/T)_S^2 \over (\delta T/T)_T^2 } = { 2
\over 25} {\alpha -1  \over \alpha} \tanh^{-2} (\psi). & (18d)} $$

The shape of $V(\psi)$ depends  on the value of $\alpha$ considered.
For $ \alpha <0$ $ (n < 1)$ it has a maximum at $\phi = 0$ and
decreases monotonically on both sides, going to $0$ for $\phi \to  \pm
\infty$. The expansion  of the universe is inflationary for all the
$\phi$ values; for large $\phi$ the potential looks as an exponential,
as in power--law inflation.

In the $\alpha >1$ case there are three subcases. For $\alpha < 1.5$
the potential has a minimum at $\phi = 0$,  two maxima, one at each
side, and it goes to $- \infty$ for $ \phi \to \pm \infty$.  For $1.5
< \alpha < 3$ the minimum is at $\phi = 0$, there are no maxima, and
$V(\phi) \to  + \infty$ for $\phi  \to \pm \infty$. Finally for
$\alpha > 3$, $\phi = 0$ is a maximum, there is a minimum at each
side and $V(\phi) \to + \infty$ for $\phi \to  \pm \infty$. The
inflation condition $\ddot a > 0$  indicates that inflation happens
only while $\phi$ evolves towards $\phi = 0$ for $\vert\phi\vert$
smaller than a given value and that it takes infinite e-foldings to
reach $\phi = 0$. Finally we see that the gravitational waves
contribution to the CMBR may dominate, but only for values of $\phi$
close to the one in which inflation starts; this can last for few
e-foldings, as equation (17) holds also in this case.

\smallskip \noindent Case $B<0$
$$\eqalignno{ {\rm a)} & \ \ \ \ V(\psi) = {(-B)^{{1 \over \alpha -
1}} \over \kappa^2 ( \alpha - 1)} \sinh ^{2(2 - \alpha)/(\alpha - 1)}
(\psi) \left(- \alpha + (2 \alpha - 3) \sinh^2 (\psi)\right)& (19a)
\cr
{\rm b)} & \ \ \ \  \tanh^2 (\psi) >
\alpha / (\alpha - 1)  \to \ddot a > 0 & (19b)\cr
{\rm c)} & \ \ \ \ \ln \left({a_y \over a_x }\right) = {1 \over 2
\alpha} \ln \left({\cosh^2 (\psi_x) \over \cosh ^2 (\psi_y)} \right)
&(19c) \cr
{\rm d)} & \ \ \ \ {(\delta T/T)_S^2 \over (\delta T/T)_T^2 } = { 2
\over 25} {\alpha -1  \over \alpha} \tanh^2 (\psi). & (19d)}$$

Condition b) indicates that there is no inflationary solution for
$\alpha > 1$. In the case $\alpha < 0$ the potential goes to $ -
\infty$ for $ \vert\phi\vert \to 0$, has two positive maxima and goes
to 0 for $\vert\phi\vert \to \infty$. Inflation only happens for large
values of $\phi$ while it evolves to $\infty$. In the large-$\phi$
limit this case reduces to the exponential potential case.

The two remaining cases, $\alpha = 1$ and $\alpha = 0$, are solved in
the same way, using equations (8) and (9) instead of equation (6).

\smallskip \noindent $\alpha = 1$ case

$$\eqalignno{{\rm a)} &  \ \ \ \ V(\phi) = {H_0 \over \kappa^2} \left(
3 - {\kappa^2 \phi^2 \over 2} \right ) \exp \left (- {\kappa^2 \over 2
A a_0} + { \kappa^2 \phi^2 \over 2} \right) & (20a)\cr
{\rm b)} & \ \ \ \ \vert \phi \vert < {\sqrt {2} \over \kappa} \to
\ddot a > 0 & (20b)\cr
{\rm c)} & \ \ \ \ \ln \left( {a_y \over a _x} \right) =
\ln \left( {\phi_y \over \phi _x} \right) & (20c)\cr
{\rm d)} & \ \ \ \ {(\delta T/T)_S^2 \over (\delta T/T)_T^2 } = { 4
\over 25} {1\over \kappa^2 \phi^2}. & (20d) }$$
The potential has a minimum at
$\phi= 0$, a maximum at each side and it goes to $- \infty$ for $\vert
\phi \vert \to \infty$. Inflation happens while $\phi$ rolls to the
minimum in a region between the two maxima. The gravitational wave
contribution is dominant only for the largest values of $\vert \phi
\vert $ satisfying the condition for inflation, and for ln (25/2)/2
 e-foldings.

\smallskip \noindent $\alpha = 0$ case

\smallskip The case of an exact scale-invariant spectrum  is given
by:
$$\eqalignno{ {\rm a)} & \ \ \ \ V(\phi) = { 64 A \over \kappa^4
\phi^2} \left( {3 \over \kappa^2} - { 32 \over \kappa^4 \phi^2}
\right) & (21a) \cr
{\rm b)} & \ \ \ \ \kappa \vert \phi \vert > 4 \sqrt
{2} \to \ddot a > 0 & (21b) \cr
{\rm c)} & \ \ \ \ \ln \left( {a_y \over a _x} \right) = {\kappa^2
\over 64} \left(\phi_y^2 - \phi_x^2 \right)& (21c) \cr
{\rm d)} & \ \ \ \ {(\delta T/T)_S^2 \over (\delta
T/T)_T^2 } = { \kappa^2 \psi^2 \over 400}. & (21d)} $$
It corresponds to the so-called Intermediate Inflation (Barrow and
Liddle 1993). The potential goes to $- \infty$ for $\vert \phi \vert
\to 0$, has two positive maxima and goes to $0$ for $ \vert \phi \vert
\to \infty$. Inflation occurs while $\phi$ goes $\to \pm \infty$ for
large $\vert \phi \vert $ values. The gravitational wave contribution
to the CMBR can dominate only for the smaller values of $\vert \phi
\vert $ compatible with inflation and for 23/4 e-foldings.
\medskip
\noindent
{\bf 3. Working models}
\medskip

 The exact solutions with $\alpha > 0$  presented in the previous
section illustrate the fact that it is possible for a minimally
coupled inflaton field to produce density perturbations with spectral
index $n>1$, provided that its potential has a particular shape. The
main characteristic is that the potential minimum has a non-vanishing
energy density, which corresponds to an effective cosmological
constant. Some simple potentials with this characteristic originate
(non-monotonically) increasing power spectra. For example $V(\phi) =
V_0 + f \phi^n$ gives rise to a spectrum increasing with $k$ for
wavelengths that left the Hubble radius when $f \phi^n < V_0 $, and
decreasing with $k$ in the opposite case.

An obvious problem with these inflationary scenarios is how to produce
the reheating. There are two main ways in which reheating is achieved
in inflationary models: through rapid oscillations of the inflaton
field around the potential minimum or through a first-order phase
transition with bubble production. The first one is associated for
example with a polynomial potential inflation model, while the second
one is associated with some extended inflation models. In the cases
discussed above, where an inflaton field rolls to a potential minimum
with large cosmological constant, none of them works. Due to the large
friction term in the Klein--Gordon equation, the scalar field does not
oscillate around the minimum but approaches it with  an
ever-decreasing velocity, taking infinite e-foldings to reach it;
thus the oscillation mechanism cannot operate. The first-order phase
transition can work, in principle, provided that the cosmological
constant corresponds to the false vacuum energy density of a second
scalar field, which at a given time makes the phase transition to its
true vacuum by bubble nucleation. However the problem with this picture
is that, as the inflationary expansion of the universe is close to
exponential, the graceful exit problem of the old inflationary
scenario appears (a slower expansion as in extended inflation would
avoid it). A possible way out of it (Linde 1991a, Adams and Freese
1991) may be to introduce a coupling between the fields, such that the
tunnelling probability is determined by the dynamics of the other
field. This reheating problem can nevertheless be avoided in a new
class of models, the ``hybrid inflation'' ones, recently proposed by
Linde (1991b,1993). In these models inflation ends in the following
way: the slow roll of the inflaton field at a given time triggers a
second-order phase transition of a second field, whose false vacuum
energy density is responsible for the cosmological constant term. The
field rolls down to its potential minimum and oscillates around it. A
particular potential in which this scenario can be realized is (Linde
1991b,1993)\footnote{$^1$}{The fact that $ n > 1$ in this model was also
noted by Liddle and Lyth 1993. After completing this work a revised
version of Linde 1993 appeared where this issue is also discussed.
The results are fully consistent.}
$$V(\phi,\sigma) = {1 \over 4\lambda} (M^2-\lambda\sigma^2)^2 +{m^2
\over 2}\phi^2 +{g^2 \over 2}\phi^2\sigma^2. \eqno (22)  $$
For $ \phi > \phi_c \equiv M/g$, $\sigma$ is in its false vacuum at
$\sigma = 0$. The potential for $\phi$ in this regime is given by $
V(\phi) = M^4 /4\lambda + m^2 \phi^2 /2$. When the cosmological
constant term dominates, this gives rise to an inflationary phase with
a power spectrum $P_\zeta (k)$ increasing with $k$. When $\phi <
\phi_c$, $\sigma$ rolls down to its minimum and the universe reheats.
The constraints put on the model parameters in order that it works are
discussed by Linde (1993). In order to have inflation for $\phi >
\phi_c$, one needs $M^2 \gg 2m^2\lambda/g^2$, and in order that
inflation finishes soon after $\phi$ reaches $\phi_c$, one needs
$M^6\ll 3 \lambda^2 M_P^4  m^2/ 2 \pi^2$. We can now look at the
resulting spectral index for the perturbations $\alpha \equiv d \ln
P_\zeta^{1/2}/ d \ln k$.

The Klein--Gordon equation for the inflaton can be written as
$$\phi'' + 3 \phi' + {12 \lambda m^2 \over \kappa^2 M^4} \phi =
0,\eqno (23)$$
where from now on a prime denotes derivative with respect to $\ln a$.
For $4 \lambda m^2 / \kappa^2 M^4 \ll 1$, the solution can be
written as
$$\phi = \phi_0 \left({a \over a_0}\right)^{-4 \lambda m^2/
\kappa^2 M^4} .\eqno (24)$$
The power spectrum $P_\zeta^{1/2} \propto H / \phi'$, and it is a good
approximation to take $d \ln k = d \ln aH \simeq d \ln a$. Thus
$$\alpha \simeq \left( \ln {H \over \phi'}\right)'= {H' \over H}-
{\phi'' \over \phi'} \simeq {4 \lambda m^2 \over \kappa^2 M^4} .\eqno
(25)$$
The spectral index is positive, as anticipated. To be consistent with
the approximations done, this result holds only for small values of
$\alpha$. This presents no problems for values of $n$ close to $1$,
such as those required to explain the large voids observed ($n \simeq
1.25$, Piran {\it et al.} 1993), which correspond to $\alpha \simeq
0.12$, consistent with the allowed range of parameters. This value
fixes the relation among $m^2$ and $M^4$ through equation (25).

The amplitude of the perturbations that left the horizon around 60
e-foldings before the end of inflation (the ones relevant for
structure formation) is given by
$$P_\zeta^{1/2} (k_{60})= {\sqrt{2\pi }g M^5 \over
\sqrt{3}\lambda^{3/2} M_P^3 m^2} \exp\left(- 60 {4 \lambda m^2 \over
\kappa^2 M^4}\right) .\eqno (26)$$
(The last factor, which was neglected in Linde 1993, has to be
included when we consider values of $ \lambda m^2 /\kappa^2 M^4 \sim
\alpha \ll 1$, but not $\ll 60$). This amplitude has to be $\sim 3
\times 10^{-5}$ in order to agree with the COBE results. This,
combined with equation (25), finally gives the values of $m$ and $M$:
$$M\simeq {\pi \sqrt{\lambda}\over g} 10^{-3} M_P,$$
$$m \simeq\sqrt{\pi\over  5}{\pi^2\lambda/g^2} 10^{-6} M_P.$$

For values of the coupling constants $\lambda$ and $g$ of order $1$
these values are consistent with the above constraints. Thus, this
provides a working model in which a power spectrum $P_\zeta^{1/2}$
increasing with $k$ can arise and a successful reheating mechanism can
be implemented to end inflation.
\medskip
\noindent
{\bf 4. Discussion and conclusions}
\medskip
The solutions obtained in section 2 illustrate the fact that it is
possible for a single minimally coupled inflaton field to produce
density perturbations with a spectrum $k^\alpha$ for any value of
$\alpha =$ const. The decreasing power spectrum ($\alpha < 0$) is the
known output of power-law inflation. Usually, spectra with
non-constant negative $\alpha$ are the results of the inflationary
potentials considered in the literature. The increasing spectra
($\alpha > 0$) are less frequent. They arise in models in which $\vert
\dot \phi \vert $ decreases more rapidly than $H^2$. This condition is
realized in the cases discussed while the inflaton rolls down to a
potential minimum with a non-vanishing potential energy (i.e. a
cosmological constant-like piece). Mathematically it is difficult to
write down a general condition for the potential to ensure that the
power spectrum increases with $k$, but it is possible to obtain the
condition in the slow-rolling approximation ($\ddot \phi \ll 3H \dot
\phi$, $V_\phi$ and $\dot \phi ^2 /2 \ll V$), that is $V V_{\phi
\phi}/V_\phi^2 > 3/2$. An inflaton field with a potential of this type
will generically have problems to reheat the universe. As we have
discussed in the last section, a possible way to circumvent this is
given by the ``hybrid inflation" models, where the coupling of the
inflaton with a second field at a given time triggers a second-order
phase transition of this field, which reheats the universe.

Another interesting issue to discuss is the contribution of the
gravitational waves to the CMBR. In all the solutions with $\alpha = $
const$ > 0$ the tensor mode can dominate over the scalar one only for
a few e-foldings as the relative contribution always decreases with
time. The reason is easy to understand, looking at equations (1) and
(17). To have a spectrum increasing with $k$, $H / \vert \dot \phi
\vert $ must increase with time.  As $H$ decreases with time, $H/
\vert \dot \phi \vert$ necessarily increases with time and thus,
according to equation (17), the ratio of the scalar to tensor mode
contributions increases with time. We have seen that the maximum
number of e-foldings, since the beginning of inflation, during which
the gravitational waves can dominate is

$$  N_{GW} = {1 \over 2 \alpha} \ln \left( 1 + {23 \over 2 } \alpha
\right). \eqno (27) $$
Thus, if gravitational waves contribute significantly to originate the
CMBR anisotropies at large angular scales, the corresponding
wavelength  should have left the Hubble radius not very far after the
beginning of inflation. This corresponds to a choice of the end of the
inflationary expansion in a specific range of values, which makes
the hypothesis unattractive. The more general prediction of these
models is a negligible gravitational wave contribution to $\Delta
T/T$.

The results presented in this work show that power spectra with $n>1$
can be easily originated in inflationary models. The astrophysical
interest in these models comes from the necessity of simultaneously
accounting for both the right amount of ultra large-scale power
required to fit CMBR anisotropies and that required to give rise to
bulk motions and structures on $\sim 50~h^{-1}$ Mpc scale. A wider
analysis of the astrophysical implications of this kind of spectra and
a comparison with the available observations are needed to test the
model further.
\bigskip
The work of S. Mollerach is supported by a grant from the Commission of
European Communities (Human capital and mobility programme). F. Lucchin
and S. Matarrese acknowledge financial support from Italian M.U.R.S.T..
\vfill
\eject
\null
\noindent
{\bf References}
\bigskip

\noindent Adams F. C. \& Freese K. 1991, Phys. Rev. D 43, 353

\noindent Adams F. C., Bond J. R., Freese K., Frieman J. A. and Olinto
A. V. 1992, Phys. Rev. D 47, 426

\noindent Barrow J. D. \& Liddle A. R. 1993, Phys. Rev. D 47,
R5219

\noindent Bunn E., White M., Srednicki M. \& Scott D. 1993, preprint

\noindent Carr B. J. \& Lidsey J. E. 1993, Phys. Rev. D 48,
543

\noindent Cen R. {\it et al.} 1992, ApJ Lett. 399, L11

\noindent Davis M. \& Peebles P. J. E. 1983, ApJ 267, 465

\noindent Davis M., Summers F. J. \& Schlegel D. 1992, Nature
359, 393

\noindent Davis R. L., Hodges H. M., Smoot G. F., Steinhardt P. J. \&
Turner M. S. 1992, Phys. Rev. Lett. 69, 1856

\noindent De Lapparent V., Geller M. J. \& Huchra J. P. 1986, ApJ
Lett. 302, L1

\noindent Dodelson S. \& Jubas J. M. 1993, Phys. Rev. Lett. 70,
2224.

\noindent Efstathiou G., Sutherland W. J. \& Maddox S. J. 1990,
Nature 348, 705

\noindent Gaier T. {\it et al.} 1992, ApJ. Lett. 398, L1

\noindent Geller M. J. \& Huchra J. P. 1989, Science 246, 897

\noindent Gorski K. M., Stompor R. \& Juszkiewicz R.  1993, ApJ Lett.
410, L1

\noindent Klypin A., Holtzman J., Primack J. R. \& Reg\"os E. (1992),
preprint

\noindent Kofman L., Gnedin N. \& Bahcall N. A. 1993, ApJ 413,
1

\noindent Lidsey J. E. \& Coles P. 1992, MNRAS 258, 57

\noindent Liddle A. R. \& Lyth D. H. 1992, Phys. Lett. {\bf B291}, 391

\noindent Liddle A. R. \& Lyth D. H. 1993, preprint

\noindent Linde A. 1991a, Phys. Lett. B 249, 18

\noindent Linde A. 1991b, Phys. Lett. B 259, 38

\noindent Linde A. 1993, preprint

\noindent Lucchin F. \& Matarrese S. 1985, Phys. Rev. D 32, 1316

\noindent Lucchin F., Matarrese S. \& Mollerach S. 1992, ApJ Lett.
401, L49

\noindent Piran T., Lecar M., Goldwirth D. S., Nicolaci da Costa L. \&
Blumenthal G. R. 1993, preprint

\noindent Salopek D. S., Phys. Rev. Lett. {\bf 69}, 3602

\noindent Schuster J. {\it et al.} 1993, ApJ Lett. 412, L47

\noindent Smoot G. F. {\it et al.} 1992, ApJ Lett. 369, L1

\noindent Souradeep T. and Sahni V. 1992, Mod. Phys. Lett. {\bf A7},
3541

\noindent Tormen G., Moscardini L., Lucchin F. \& Matarrese S. 1993,
ApJ 411, 16

\noindent Turner M. 1991, Physica Scripta 36, 167

\noindent Vittorio N., Matarrese S. \& Lucchin F. 1988, ApJ 328,
69

\end